\documentstyle[12pt]{article}
\newcommand{\be}{\begin{equation}}
\newcommand{\ee}{\end{equation}}
\newcommand{\bea}{\begin{equation}\begin{array}{lcl}}
\newcommand{\eea}{\end{array}\end{equation}}
\newcommand{\ba}{\begin{array}}
\newcommand{\ea}{\end{array}}

\newcommand{\N}{\rm N \kern -2.6mm I}

\newcommand{\norsl}{\normalsize\sl}
\newcommand{\norsc}{\normalsize\sc}
\begin{document}

\textheight 22cm
\voffset  -1cm
\begin{titlepage}

\title{\Large\bf {Finite Unified Models}}

\author{
\norsc  D. Kapetanakis\thanks{Supported by a C.E.C. fellowship
(ERB4001GT910195).}, M. Mondragon\thanks{Supported by an A. von Humboldt
fellowship.}  \\
\norsl  Physik Department\\
\norsl  Technische  Universit\"{a}t M\"{u}nchen\\
\norsl  D-8046 Garching, Germany\\
\\
and \\
\\
\norsc G. Zoupanos \thanks{Partially supported by a C.E.C.
project (SC1-CT91-0729)}\\
\norsl  Physics Deptartment\\
\norsl  Nat. Technical University\\
\norsl  157 73 Zografou, Athens, Greece}

\date{}
\maketitle

\begin{abstract}
{\normalsize We present phenomenologically viable $SU(5)$ unified models which
are finite to all orders before the spontaneous symmetry breaking. In the case
of two models with three families the top quark mass is predicted to be 178.8
GeV.}
\end{abstract}

\begin{picture}(5,2)(-300,-550)
\put(2.3,-110){MPI--Ph/92-70}
\put(2.3,-130){TUM--TH--145/92}
\end{picture}

\thispagestyle{empty}
\end{titlepage}
\setcounter{page}{1}
\baselineskip=6mm
\renewcommand{\arraystretch}{1.7}
%%%%%%%%%%%%%%%%%%%%%%%%%%%%%%%%%%%%%%%%%%%%%%%%%%%%%%%%%%%%%%%%

\section*{\large\bf 1. Introduction}

The apparent success of unified gauge theories describing the observed
interactions is restrained by the plethora of arbitrary parameters that one has
to introduce by hand. In particular, in the electroweak standard model
\cite{SM}, which is indeed a very successful theory, one has to fit more than
twenty parameters if neutrinos are massive or eighteen if they are massless.
This is a clear disadvantage as far as the predictivity of the theory is
concerned. Grand Unified Theories (GUTs) \cite{SU5,GUTS}
 are doing better in this
respect since they can provide predictions for parameters such as
$\sin^2\theta _W$ and fermion mass ratios, which are free parameters in the
electroweak standard model. In turn, GUTs can be tested and possibly could be
ruled out, as for instance is the case of the minimal $SU(5)$ model \cite{exp}.

There exists another principle that certainly points to the direction of
further reduction of the free parameters of a gauge theory, namely, the
requirement of finiteness. Moreover, the principle of finiteness goes very
deeply to the heart of quantum field
 theories, supporting strongly the hope that
the ultimum theory does not need infinite renormalizations. Although the latter
are perfectly legitimate in quantum field theory they still give the feeling
that divergences are ``hidden under the carpet'' \cite{44}. It is not
accidental that supersymmetric gauge theories have been so widely explored
during
the last decade in spite of the lack of any experimental evidence of
supersymmetry. The clear motivation for the explosion of interest is due to the
absence of quadratic divergences in these theories which guarantees their
naturalness.

There have been made many attempts to obtain finite quantum field theories in
four dimensions. For general theories such searches are usually limited to one
loop approximation \cite{55}. Besides, there is a strong indication
that only supersymmetric gauge theories can be completely free from ultraviolet
divergences \cite{55}. A very interesting fact is that the one loop finiteness
conditions on $N=1$ supersymmetric theories automatically ensure also two-loop
finiteness \cite{PW}. Last but not least, there have been given simple criteria
\cite{LPS,alloop,al} which ensure ``all orders finiteness'' in
 the sense of vanishing $\beta$-functions.

A complete classification of chiral $N=1$ supersymmetric theories with a simple
gauge group that satisfy the one-loop finiteness conditions has been done in
refs. \cite{HPS,JZ}. There appear to exist only a few possibilities that have
a chance to develop to realistic models. Here we examine to which extent these
models can be made realistic, imposing in addition the requirement of all
orders finiteness in the sense of ref. \cite{LPS}. We find interesting
solutions to this problem. Furthermore, in the case of the models involving
three generations a heavy top quark naturally
emerges, a feature which seems
to be characteristic of this class of models.

\section*{\large\bf 2. Finite $N=1$ Supersymmetric Gauge Theories}

In order to discuss in detail the finiteness conditions and their implications,
let us consider a chiral, anomaly free, globally supersymmetric $N=1$ gauge
theory with gauge group $G$. The superpotential of such a theory is given by:
\be
W=a_i \phi _i +{1\over 2} m_{ij}\phi _i \phi _j+
 {1\over 6} C_{ijk}\phi _i \phi _j \phi _k \,,
\label{supot}
\ee
where $a_i$, $m_{ij}$ and $C_{ijk}$ are gauge invariant tensors and the matter
fields $\phi _i$ transform according to an irreducible representation $R_i$
 of the
gauge group $G$.

The necessary and sufficient conditions for finiteness at one-loop level are
the following:
\begin{itemize}
\item One-loop finiteness of the gauge fields self-energy which requires:
\be
\sum _i \ell (R_i) = 3 C_2(G) \,,
\label{1st}
\ee
where $\ell (R_i)$ is the Dynkin index of $R_i$ \cite{slansky}
and $C_2(G)$ is the quadratic
Casimir operator of the adjoint representation of the gauge group $G$.

\item One-loop finiteness of the chiral superfields self-energy.
 In terms of the
cubic couplings $C_{ijk}$ appearing in the superpotential given in
eq.~(\ref{supot}), referred to as Yukawa couplings, this condition requires:
\be
C^{ikl} C_{jkl} = 2\delta ^i_j g^2  C_2(R_i)\,,
\label{2nd}
\ee
where $g$ is the gauge coupling constant, $C_2(R_i)$ is the quadratic
Casimir of the representation $R_i$, and $C^{ijk}= (C_{ijk})^*$. Note that
condition \ref{2nd} forbids the presence of singlets with nonzero coupling.
Furthermore, it requires that $C^{ikl} C_{jkl}$ is diagonal in its two free
indices.
\end{itemize}

Therefore, the finiteness conditions given in eqs.~(\ref{1st}) and (\ref{2nd}),
 which express the
vanishing of the one-loop anomalous dimensions of the gauge and matter
couplings respectively, restrict considerably the choices of the
representations $R_i$s for a given group $G$ as well as their Yukawa couplings
appearing in the superpotential, eq.~(\ref{supot}).
 On the other hand due to the
non-renormalization theorem \cite{wzif}, which relates the renormalization of
 $a_i$, $m_{ij}$ and $C_{ijk}$ to that of the $\phi _i$, the finiteness
 conditions do not restrict the form of  $a_i$ and $m_{ij}$.

An important consequence of the finiteness conditions is that supersymmetry
most probably can only be broken by the addition of soft breaking terms.
Specifically, due to the exclusion of singlets according to eq.~(\ref{2nd})
the F-type \cite{f} spontaneous supersymmetry
 breaking terms are incompatible with
finiteness. Also, the D-type \cite{d} spontaneous breaking is also ruled out
since it requires the existence of a $U(1)$ gauge group which
in turn is incompatible with eq.~(\ref{1st}).
In choosing to break supersymmetry by the addition of soft terms one
should be aware of the fact that one-loop finiteness imposes extra conditions
on this sector  of the theory \cite{jmy}.

A very interesting result proved in ref.~\cite{PW} is that the one-loop
finiteness conditions (\ref{1st}), (\ref{2nd}) are necessary and sufficient for
finiteness at two-loop level. Even more interesting is the theorem proved in
ref.~\cite{LPS}. The theorem states that if a supersymmetric
gauge theory with simple gauge group is free from gauge anomalies, obeys
eq.~(\ref{1st}), and there exist solutions to eq.~(\ref{2nd}) of the form
\be
C_{ijk}=\rho _{ijk} g \,,
\label{sol}
\ee
where $\rho _{ijk}$ is a complex number, which are isolated and non-degenerate,
then each of these solutions can
be uniquely extended to a formal power series of $g$ \cite{Zim},
 giving a theory which
depends on a single coupling $g$, with a $\beta$-function vanishing to all
orders.

\section*{\large\bf 3. Finite Unified Models based on SU(5)}

An inspection on the tables of refs.~\cite{HPS,JZ} immediately shows
the difficulties encountered in constructing phenomenologically viable finite
unified theories (FUTs) already at the one- or equivalently two-loop level. In
particular, using $SU(5)$ as gauge group there exist only two candidate models
which can accommodate three fermion generations and they contain the chiral
multiplets 5, $\bar 5$, 10, $\overline {10}$, 24 with multiplicities
(6,9,4,1,0) and (4,7,3,0,1) respectively. In addition, there exists another
model
based on $SU(5)$ gauge group which can accommodate five fermion generations and
contains the same chiral multiplets as the two previous with multiplicities
(5,10,5,0,0). Out of these three models only the second one contains a 24-plet
which can be used for the spontaneous symmetry breaking of $SU(5)$ down to the
standard model $SU(3)\times SU(2)\times U(1)$. For the other two models one has
to incorporate another way such as the Wilson flux breaking mechanism
\cite{wfb} in order to achieve the required superstrong spontaneous symmetry
breaking of the $SU(5)$ gauge group.

In the following we will consider in more detail the three family models.

\subsection*{\sl A. $N=1$, $SU(5)$ model with three fermion families and
without adjoint Higgs}

 The particle content of this model consists of the following supermultiplets
 represented by their transformation properties under $SU(5)$:
 three ($\bar 5$+10), which are identified with the three supermultiplets
 describing the fermion families, six ($5+{\bar 5}$) which are considered  as
 Higgs supermultiplets, and one ($10 + {\overline {10}}$) which are considered
also
 as scalar supermultiplets.

The first finiteness condition given  in eq.~(\ref{1st}) is automatically
satisfied in the present model given that this was one of the selection rules
for the models appearing in refs.~\cite{HPS,JZ}. In order to satisfy  the
second condition given in eq.~(\ref{2nd}) we
have to consider the superpotential.
The most general $SU(5)$ invariant, $N=1$ cubic superpotential with the above
particle content has the form:
\bea
W &=& {1\over 2} g_{ija} 10_i 10 _j  H_a + g_{ia} 10_i N H_a
+ {\bar g}_{ija} 10_i {\bar 5}_j {\bar H}_a
+{1\over 2} {g^\prime}_{ijk} 10_i {\bar 5}_j {\bar 5}_k \\
&+&
{1\over 2}f_{a b} N{\bar H}_a {\bar H}_b
+
{1\over 2}{\bar f}_{a b} {\bar N} H_a H_b
+
{1\over 2}h_a N N H_a
+
{1\over 2}{\bar h}_a {\bar N}{\bar N} {\bar H}_{a}\\
&+&
{1\over 2}q_{ia b} 10_i {\bar H}_a {\bar H}_b
+ p_{i a} N {\bar 5}_i {\bar H}_a +{1\over 2}t_{ij} N{\bar 5}_i {\bar 5}_j \,,
\label{superpot}
\eea
where $i,j,k=1,\dots ,3$ and $a ,b =1,\dots ,6$ and we have suppressed
the $SU(5)$ indices.
 $10_i$ and ${\bar 5}_i$
are the usual three generations. The six ($5+{\bar 5}$) Higgses are
denoted by $H_a$, ${\bar H}_a$,  while the  scalar field
 belonging to the ($10 + {\overline {10}}$) representation by $N+ {\bar N}$.

Then, eq.~(\ref{2nd}) imposes the following relations among the Yukawa and
gauge couplings:
\bea
H:\quad 3g^{ija}g_{ijb} + 6 g^{ia}g_{ib} + 4 f^{ca}f_{cb} + 3h^a h_b &=&
\delta ^a_b {24 \over 5} g^2\,, \\
{\bar 5}:\quad 4{\bar g} ^{ila} {\bar g}_{ima} + 4 {g'} ^{ilk} {g'}_{imk} +4
t^{lj}t_{mj} +
p^{la}p_{ma} &=& \delta ^l_m{24 \over 5} g^2\,,\\
{\bar H}:\quad 4{\bar g} ^{ija} {\bar g}_{ijb} + 4f^{ca}f_{cb} + 3{\bar
h}^a{\bar h}_b +
4q^{ica}q_{icb} + 4p^{ia}p_{ib} &=& \delta ^a_b{24 \over 5} g^2\,,\\
N:\quad 3g^{ia}g_{ia} + f^{cb}f_{cb}+ 3h^ah_a + 2p^{ia}p_{ia} +t^{ij}t_{ij} &=&
  {36 \over 5} g^2\,,\\
{\bar N}:\quad {\bar f}^{ab}{\bar f}_{ab} + 3{\bar h}^a{\bar h}_a &=& {36 \over
5} g^2\,,\\
10:\quad 3g^{lki}g_{mki}+2{\bar g} ^{lki} {\bar g}_{mki} + 3 g^{la}g_{ma} +
{g'}^{ljk}{g'}_{mjk} + q^{lab}q_{mab}
&=&\delta ^l_m{36 \over 5} g^2\,.
\label{Yg1}
\eea

As it was already emphasized in sect.~2
the fulfillment of eqs.~(\ref{1st}) and (\ref{2nd}) is necessary and sufficient
to guarantee the one-loop as well the two loop finiteness of the theory
\cite{PW}.
Nevertheless, in order to achieve all-loop finiteness one has to do more
\cite{LPS}. Specifically, one has to find a solution of eq.~(\ref{2nd}) which
is isolated and non-degenerate. This is a far from trivial
problem given that eq.(\ref{2nd}) have infinitely many  solutions that can be
parametrized by continuous parameters (look for example
refs.~\cite{model1,model}).

Our strategy to find a unique and phenomenologically interesting solution to
eq.~(\ref{2nd}) is to impose on the model additional symmetries on  top of
the $SU(5)$ gauge invariance and $N=1$ global supersymmetry. Next recall that
the terms of lower dimensions such as mass terms are not restricted by the
finiteness requirement. We use this freedom to make the model
phenomenologically viable. As a result we found a solution to all-loop
finiteness problem with very interesting phenomenological predictions. In
particular the top quark mass is predicted. The method can be generalized in a
straightforward way in order to take into account all light fermion masses and
mixing angles \cite{kmz}.
\begin{table}[t]
\caption{The charges of the $Z_7\times Z_3$ symmetry}
\label{tbl}
$$
\begin{tabular}{|c|c|c|c|c|c|c|c|c|c|c|c|c|c|}
 \hline
  & $10_1$  & $10_2$ & $10_3$ & $\bar 5_1$ & $\bar 5_2$ &  $\bar 5_3$& $H_1$ &
 $H_2$ & $H_3$ & $H_4$ &
 $H_5$ & $H_6$& $N$ \\ \hline
$Z_7$ &1 &  2    &  4 &   4   &   1   &   2   & 5&   3  &  6  & 0 &
 0& 0&0   \\ \hline
$Z_3$ &1 &2     &0    & 0 &0 &0 &1 & 2 & 0 & 0 &
 1 & 2& 0 \\ \hline
\end{tabular}
$$
\end{table}
Specifically, we impose the $Z_7\times Z_3$
 discrete symmetry given in
table \ref{tbl}, together with a multiplicative Q-parity
under which the $10_i$
and $\bar 5 _i$ describing the fermion supermultiplets are odd, while all the
other superfields are even. In this way the number of terms that are permitted
to appear in the superpotential is severely restricted. Only terms with Yukawa
couplings $g_{iii}$, ${\bar g}_{iii}$, $f_{44}$, $f_{56}$, $f_{65}$,
 ${\bar f}_{44}$, ${\bar f}_{56}$, ${\bar f}_{65}$, $h_4$, and $\bar h_4$
 survive.

We then find the following unique solution to eqs.~(\ref{Yg1}),
\bea
g_{111}^2=g_{222}^2=g_{333}^2 = {8\over 5} g^2 \,,\\
{\bar g}_{111}^2={\bar g}_{222}^2= {\bar g}_{333}^2= {6\over 5} g^2 \,,\\
f_{44}^2 = 0 ;~~ f_{56}^2 = f_{65}^2={6\over 5}g^2\,, \\
\bar f_{44}^2 = 0 ;~~\bar f_{56}^2 =\bar f_{65}^2 = {6\over 5}g^2  \,,\\
h_4^2 ={8\over 5}g^2;~~\bar h_4^2 ={8\over 5}g^2\,.
\label{solu}
\eea
\vfil\break
\noindent
The uniqueness \footnote{The phase arbitrariness of eqs.~(\ref{solu}) is not
crucial, since it can be absorbed by using a specific renormalization
 scheme \cite{LPS}.}
 of this solution guarantees the all-loop finiteness.

One might wonder if this model could result from some more fundamental theory
and, in turn, if there is some justification for its symmetries. It seems that
there exist very suggestive hints that the model under consideration belongs to
a class of models obtained from superstring compactification over certain
Calabi--Yau (CY) manifolds. More specifically, Witten \cite{wtt} has shown
 that it is possible to
construct stable, irreducible, and holomorphic $SU(5)$ or $SU(4)$ vector
bundles over CY manifolds.
 Then one can start from the heterotic superstring
 with gauge group  $E_8\times E_8^\prime$ and obtain an  $SU(5)$
or $SO(10)$ $N=1$ supersymmetric theory at four dimensions,
 by embedding the structure group
of the bundle ($SU(5)$ or $SU(4)$) in $E_8$ ($E_8^\prime$ is considered as
hidden).
It is worth noting that claims that such configurations
 are generically unstable \cite{dsww}
due to non-perturbative effects appeared unjustified in particular cases.
Furthermore the conditions under which a stable configuration emerges are given
in ref.~\cite{distler}. It turns out that the spectrum of a $N=1$, $SU(5)$
gauge theory resulting from a CY compactification is
generally of the form  $m(10)+n({\overline 5})+\delta (10+{\overline {10}})+
\epsilon (5+{\bar 5})$, where $m$, $n$, $\delta$, and $\epsilon$ are
 topological numbers of the CY manifold \cite{greene}.
Therefore, it is not inconceivable to imagine how a model
 like the one considered here could come from superstring compactification.

Furthermore, since in the present model
 we are interested in applying the Wilson
flux breaking mechanism, we, naturally, assume that the CY which is going to be
used should admit a freely acting discrete group $F$.
Then the light fields will be the ones
which are invariant under $T\oplus F$, where $T$ is the homomorphism of $F$ in
the gauge group.

Therefore, we are led to assume the existence of a CY with a stable,
irreducible, and holomorphic $SU(5)$ bundle over it,
admitting a freely acting discrete
group $F$. Moreover, the topological numbers of this manifold after division
with $F$ are given by  $m=n=3$, $\delta =1$, and $\epsilon =6$. Let us comment
here that the discrete symmetries used above in order to reduce the number of
the Yukawa couplings should be respected by this CY manifold.

The present model clearly belongs to the class of models considered in
ref.~\cite{greene}. For instance, suppose that $F$ is a $Z_3$ which is embedded
in a $T=Z_3$ identified with a discrete subgroup of the $U(1)$ appearing in the
decomposition
\bea
SU(5) &\supset& SU(3)\times SU(2)\times U(1)\\
10 &=& (1,1)(6)+({\bar 3},1)(-4)+(3,2)(1)\\
{\bar 5} &=&  (1,2)(-3)+({\bar 3},1)(2)\,.
\label{dec}
\eea
Next recall that the gauge symmetries surviving after applying the Wilson flux
breaking mechanism are those that commute with $T$. Then it is clear that the
SU(5) gauge symmetry of the model at hand breaks down to the standard model.
One can go further and consult the tables of ref.~\cite{greene} in order to
attribute appropriate transformation properties to the various scalar
multiplets, such as to make the model phenomenologically viable. As an example,
consider that the scalar multiplets are invariant under the action of $F$,
while they transform under the action of $T$ according to $\exp (y\pi)$ where
$y$ is the hypercharge in eq.(\ref{dec}). Then one can easily see that only the
$(1,2)(-3)$ components coming from the
$\bar 5$ and the $(1,1)(6)$ coming from the 10 remain light. All the other
components acquire superheavy masses of the order of the compactification
scale.
Therefore, in a natural way the model is provided with light Higgs doublets
that can drive the spontaneous symmetry breakdown of $SU(2)\times U(1)$ down to
$U(1)_{em}$ and, on the other hand, it is exorcised from the appearance of
light
``coloured scalars'' that would lead to fast proton decay. Note that the above
discrete symmetries do not affect the fermion supermultiplets \cite{greene}.

Having described the basic strategy to make the model
phenomenologically viable we postpone the full analysis of the various
possibilities to a future publication \cite{kmz}. For our purposes here we
assume that the discrete symmetries involved permit only the existence of a
pair of light Higgs doublets which is coupled only to the third family.
Moreover, by adding soft breaking terms we can achieve supersymmetry breaking
at
the order of the electroweak scale. Then examining the evolution of the gauge
couplings according to the renormalization group equations \cite{rge} we find
\bea
\sin ^2 \theta _W (M_Z)= 0.233\,,
 M_X = 2\cdot  10^{16}\,,
\alpha_{em}^{-1}(M_Z) = 127.9 \,, \\
\alpha _s (M_Z) = 0.120\,,
 {\rm and}~ \alpha _X = 0.0425\,,
\eea
in excellent agreement with the experimental values \cite{exp}
\bea
\sin ^2 \theta _W (M_Z)~^{exp}=0.2327\pm 0.0008\,,\\
\alpha _{em}^{-1}(M_Z)~^{exp} = 127.9\pm 0.2\,,\\
\alpha _s (M_Z) ~^{exp}= 0.118\pm 0.008\,.
\eea
Running now the renormalization group equations for the Yukawa couplings with
the above values for $\alpha_X$ and $M_X$ and initial values at $M_X$:
\be
g_t^2={8\over 5}(4\pi \alpha _X);~~ {\bar g}_b^2={\bar g}_{\tau}^2= {6\over 5}
(4\pi \alpha _X)
\ee
we find at $M_W$:
\be
m({\rm top})=178.8~ {\rm GeV}\,,
{}~~~m({\rm bottom})=3.1~ {\rm GeV}\,,~~ {\rm and}~~
m({\rm tau})=1.8~ {\rm GeV}\,.
\ee
As we can see, the model gives results for the tau and bottom masses in very
good agreement with experiment, and predicts a high value for the mass of the
top.
Notice that these values are determined by the solution (\ref{solu}) to the
finiteness conditions (\ref{Yg1}), and that although we have assumed that
only the third family becomes massive, we do not expect the results to change
considerably, since the third family terms dominate in the calculation.

\subsection*{\sl B. $N=1$, $SU(5)$ model with three fermion families and
Higgs in the adjoint}

This model has been considered before for two-loop \cite{model1,model} as well
as for all-loop finiteness \cite{al}.
The particle content
consists of the following supermultiplets:
three ($\bar 5+10$), identified with the three supermultiplets
describing the fermion families, four ($5+{\bar 5}$), and one 24 considered  as
Higgs supermultiplets.

The first finiteness condition, eq.~(\ref{1st}), is, as before
 automatically met. In
order to satisfy the second condition, eq.~(\ref{2nd}), we have to examine the
superpotential of the model. The most general $SU(5)$ invariant, $N=1$  cubic
superpotential with the above particle content is:
\bea
W &=&{1\over 2} g_{ija} 10_i 10 _j  H_a
+ {\bar g}_{ija} 10_i {\bar 5}_j {\bar H}_a
+{1\over 2}{g^\prime}_{ijk} 10_i {\bar 5}_j {\bar 5}_k \\
&+&
{1\over 2}q_{ia b} 10_i {\bar H}_a {\bar H}_b
+f_{a b} {\bar H}_a 24 H_b
+p (24)^3
+h_{i a} {\bar 5}_i 24  H_a \,,
\label{superpot2}
\eea
where $i,j,k=1,\dots ,3$ and $a ,b =1,\dots ,4$ and we have suppressed
the $SU(5)$ indices.
 The $10_i$'s and ${\bar 5}_i$'s
are the usual three generations, and 24 is the scalar superfield in the
adjoint.
 The four ($5+{\bar 5}$) Higgses are
denoted by $H_a$, ${\bar H}_a$.

Then, eq.~(\ref{2nd}) imposes the following relations among the Yukawa and
gauge couplings:
\bea
\bar H:\quad 4 {\bar g}_{ij a}{\bar g}^{ijb} + {24 \over 5}
f_{ac}f^{bc}+4q_{iac}q^{ibc} &=&
{24\over 5} g^2 \delta _a ^b \,,\\
H: \quad 3g_{ija}g^{ijb}+ {24 \over 5} f_{ca}f^{cb}+ {24 \over 5} h_{ia}h^{ib}
&=&
{24\over 5} g^2 \delta _a ^b \,,\\
\bar 5 :\quad 4 {\bar g}_{ki a}{\bar g}^{kja}+{24 \over 5}
h_{ia}h^{ja}+4{g^\prime}_{ikl}{g^\prime}^{jkl} &=&
{24\over 5} g^2 \delta _i^j\,,\\
10 :\quad2 {\bar g}_{ik a}{\bar g}^{jka}+3g_{ika}g^{jka}+q_{iab}q^{jab}
+{g^\prime}_{kli}{g^\prime}^{klj} &=&
{36\over 5} g^2 \delta _i ^j\,,\\
24:\quad f_{ab}f^{ab}+{21\over 5} pp^*+ h_{ia}h^{ia} &=& 10g^2\,.
\label{yg}
\eea
In most of the previous studies of this model no attempt
 was made to find isolated and
non-degenerate solutions. Their philosophy was rather in the opposite
direction. They have used the freedom offered by the degenerate solutions in
order to make specific ansatze that could lead to phenomenologically
acceptable predictions. Following the lines prescribed in the previous model we
impose additional symmetries on the model \footnote{See however ref.~\cite{al}
for an attempt to construct an all-loop finite model.}.
 The new symmetries imposed on this
model are again given in table \ref{tbl} for $10_i$, $\bar 5 _i$ and $H_a$ for
$a=1,\dots ,4$. The terms  in the superpotential which are invariant under the
symmetries of the model are the terms with Yukawa
couplings $g_{iii}$, ${\bar g}_{iii}$, $f_{ii}$ and $p$.

We find the following solution of eq.~(\ref{yg})
\bea
g_{111}^2=g_{222}^2=g_{333}^2 = {8\over 5} g^2 \,,,~~ {\bar g}_{111}^2={\bar
g}_{222}^2= {\bar g}_{333}^2= {6\over 5} g^2 \,,\\
f_{11}^2 = f_{22}^2=f_{33}^2=0\,,~~f_{44}^2=g^2;~~p^2={15\over 7}g^2\,.
\eea
Therefore, we are in the same situation as in ref.~\cite{model},
 i.e. each fermion family
is coupled to a different Higgs. For simplicity, as in the previous models, we
assume that only one pair of Higgs fields is light and acquires a v.e.v. which
is
coupled to the third family. This situation can easily be realised by adding
appropriate mass terms. The solution of the doublet-triplet splitting problem
in this model goes along the lines described in  ref.~\cite{model}.

\section*{\large\bf 4. Finite Models based on other gauge groups}

There exist some more FUTs that have a chance to develop into realistic models.
For instance, an inspection of the list of  refs.~\cite{HPS,JZ} suggests that
the following models are worth to be examined:
\begin{enumerate}
\item An $SO(10)$ model with particle content consisting of eight 10,  $n~16$
and $(8-n)~{\overline {16}}$ (with $5\leq n \leq 8$) supermultiplets.
 This model can accommodate an even number of fermion
families and could result from a CY compactification as it was discussed in
model {\sl A}.

\item An $E_6$ model containing $n~27$ and  $(12-n)~\overline {27}$
(with $7\leq n \leq12$)  supermultiplets which can accommodate an even number
of fermion families.

\item An $SU(6)$ model with three 6, nine $\bar 6$ and one 35 supermultiplets.
The model can describe three fermionic families, six Higgs in the fundamental,
six Higgs in the antifundamental and one Higgs in the adjoint.
\end{enumerate}

\section*{\large\bf 5. Conclusions}

We have discussed a number of one and two loop finite unified models. Emphasis
was given in the construction of $SU(5)$, $N=1$ supersymmetric models which are
finite in all orders before the spontaneous symmetry breaking.

In particular, in the case of $SU(5)$, $N=1$ supersymmetric models with three
families the top quark mass is predicted to be 178.8 GeV. We have restricted
our analysis to the case that only the third fermion family becomes massive
after the electroweak symmetry breaking. The generalization to non zero masses
for the rest fermions and mixing angles is straightforward. However, due to the
clear dominance of the third family, and in particular of the top quark mass,
our prediction is not expected to change in a noticeable way.
\begin{center}
{\large\bf Acknowledgements}
\end{center}
\noindent
The authors would like to thank H.-P. Nilles,
C. Lucchesi and M. Aschenbrenner for helpful
discussions. G.Z. would also like to thank M. Quiros and C. Savoy.

\end{document}